\documentclass[reprint,superscriptaddress,amsmath,amssymb,aps, longbibliography]{revtex4-1}
\usepackage{filecontents}

\usepackage{amsfonts}
\usepackage{amsmath}
\usepackage{amssymb}
\usepackage{graphicx}
\usepackage{dcolumn}
\usepackage{bm}

\begin{document}


\title{Thermal effects in spin torque switching of perpendicular magnetic tunnel junctions at cryogenic temperatures}

\author{L. Rehm}
\email{laura.rehm@nyu.edu}
\affiliation{Center for Quantum Phenomena, Department of Physics, New York University, New York, NY 10003, USA}
\author{G. Wolf}
\affiliation{Spin Memory Inc., Fremont, CA 94538, USA}
\author{B. Kardasz}
\affiliation{Spin Memory Inc., Fremont, CA 94538, USA}
\author{E. Cogulu}
\affiliation{Center for Quantum Phenomena, Department of Physics, New York University, New York, NY 10003, USA}
\author{Y. Chen}
\affiliation{Center for Quantum Phenomena, Department of Physics, New York University, New York, NY 10003, USA}
\author{M. Pinarbasi}
\affiliation{Spin Memory Inc., Fremont, CA 94538, USA}
\author{A. D. Kent}
\email{andy.kent@nyu.edu}
\affiliation{Center for Quantum Phenomena, Department of Physics, New York University, New York, NY 10003, USA}

\date{\today}

\begin{abstract}
Temperature plays an important role in spin torque switching of magnetic tunnel junctions causing magnetization fluctuations that decrease the switching voltage but also introduce switching errors. Here we present a systematic study of the temperature dependence of the spin torque switching probability of state-of-the-art perpendicular magnetic tunnel junction nanopillars (40 to 60~nm in diameter) from room temperature down to 4~K, sampling up to a million switching events. The junction temperature at the switching voltage---obtained from the  thermally assisted spin torque switching model---saturates at temperatures below about 75~K, showing that junction heating is significant below this temperature and that spin torque switching remains highly stochastic down to 4~K. A model of heat flow in a nanopillar junction shows this effect is associated with the reduced thermal conductivity and heat capacity of the metals in the junction.
\end{abstract}


\maketitle

\section{\label{sec:level1}Introduction}
Spintronic devices based on spin-transfer torques (STT)~\cite{Slonczewski1996, Berger1996} have attracted a great deal of interest in recent years due to their non-volatility, energy efficiency, small foot prints, fast operation and high reliability~\cite{Ohno2010,Park2012,Jan2014, Ikeda2010}.  Perpendicularly magnetized magnetic tunnel junction (pMTJ) are currently the most promising and extensively studied STT device for commercialization because of their high switching efficiency and scaling properties~\cite{KentWorledge2015,Jinnai2020}. Although commercial operating temperature are between -40 and 150$^{\circ}$C, pMTJs have also been recently explored for use as memory elements for a cryogenic computer operating at $\simeq 4$~K~\cite{Cao2019,Rehm2019,Lang2020}. Interest in these devices is associated with the need for a high density low temperature memory that can be tightly integrated with superconducting logic~\cite{Holmes2013}.

It is generally thought that decreasing the device temperature, while increasing the switching voltage, would lead to more reliable switching. This is because thermal fluctuations introduce randomness in the switching process that produces write errors and read disturbs. These are characterized by a thermal activation model for spin torque switching that is also often used to assess key device metrics, including the switching efficiency~\cite{Sun2013}, the ratio of the energy barrier to magnetization reversal to the spin-torque switching threshold. While studies of devices at and above room temperature are quite common there are few studies over a broad temperature range down to 4~K, the temperature relevant for applications in superconducting electronics, and no studies at 4~K that explore the switching probability with a million events.

\begin{figure}
\includegraphics[width=0.48\textwidth,keepaspectratio]{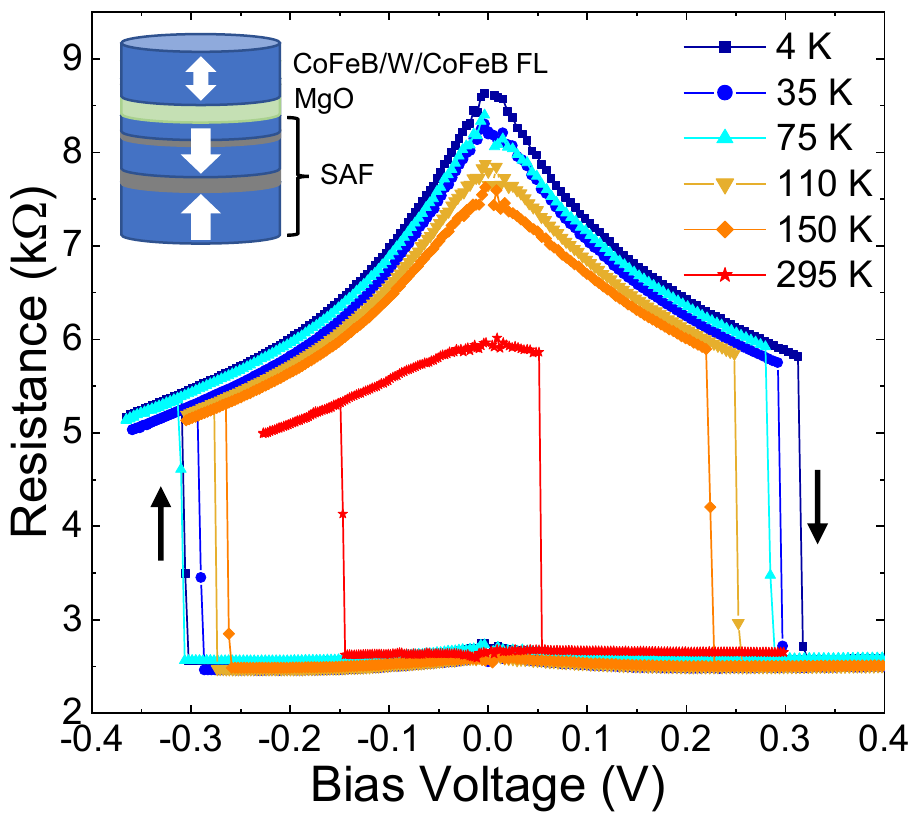} 
\caption{Quasistatic voltage-induced switching of a 40~nm pMTJ at several temperatures and zero field. The TMR ratio around zero applied voltage increases by almost a factor of two from 124\% at room temperature to 220\% at 4~K. The inset shows a schematic of the studied junctions with CoFeB/W/CoFeB composite free layer.}
\label{Fig:Fig1}
\end{figure}

In this article, we use the thermally activated spin-transfer torque switching model~\cite{Sun2000,Myers2002,Koch2004,LiZhang2004} to determine the effective switching temperature $T_\mathrm{eff}$---the sample temperature at the switching voltage---and the voltage switching threshold from room temperature down to 4~K. Two different methods are used, one based on measurements of the read disturb rate and the other on the measurements of the switching voltage versus pulse duration. These are applied to 40 to 60~nm diameter pMTJs with relatively low resistance products (RA $\approx$ 3~$\Omega\mu$m$^2$), junctions that  exhibit low energy ($<300$~fJ/switch) and fast (sub-ns) switching at 4~K as reported in~\cite{Rehm2019}.  We find that the switching temperature $T_\mathrm{eff}$ versus bath temperature saturates below about 75~K. Our findings show that junction heating is significant below this temperature and that spin torque switching remains stochastic at cryogenic temperatures.

\section{Magnetic tunnel junctions}
We investigated magnetic tunnel junctions with a composite CoFeB free layer (FL), CoFeB(1.5)/W(0.3)/CoFeB(0.8) and a CoFeB(0.9) reference layer (RL) separated by a 1~nm thick MgO tunnel barrier. Numbers in parentheses are the layer thicknesses in nm. The W layer adds perpendicular magnetic anisotropy and acts as a Boron getter during junction annealing. In contrast to room temperature devices, the FL only has one MgO interface instead of two~\cite{Kim2015,Devolder2018}, as large perpendicular anisotropy is not needed for low temperature applications. The RL is coupled to a synthetic antiferromagnet (SAF) which is composed of two Pt/Co multilayers that are antiferromagnetically coupled through a 0.8~nm thick Ru layer. The layer stack CuN(10)/SAF(9.8)/RL/MgO/FL (inset of Fig.~\ref{Fig:Fig1}) was deposited at room temperature and annealed for 25~min at 400$^\circ$C. Electron beam lithography and Ar\textsuperscript{+} ion beam milling was then used to fabricate 40, 50 and 60~nm diameter circular-shaped nanopillars.
 
We first characterized the pMTJ devices by measuring their resistance as a function of applied DC bias voltage at various temperatures between 4 and 295~K. The bias voltage is applied using a DAQ board (National Instruments PCIe-6353), which simultaneously measures the voltage drop across the tunnel junction and a resistor in series with the junction. The voltage drop across the resistor is used to determine the current flow through the junction. The measurements are performed in a cryogenic probe station where the sample stage can be heated up to 150~K. The room temperature measurements are performed while the cryostat cold head was turned off. 

Figure~\ref{Fig:Fig1} shows the junction resistance versus bias voltage and voltage-induced switching of a 40~nm pMTJ device at different temperatures in zero applied field. While the resistance in the parallel (P) state $R_\mathrm{P}$ (lower resistance branch in the figure) is almost independent of temperature and bias voltage, the resistance in the antiparallel (AP) state $R_\mathrm{AP}$ (upper branch) shows a strong temperature and bias dependence. This can be attributed to inelastic tunneling processes~\cite{Zhang1997,Shang1998,Khan2010}. In the AP state inelastic processes open additional conduction pathways as the bias is increased leading to this characteristic inverted ``V'' broken-linear response~\cite{Slonczewski2007}. Figure~\ref{Fig:Fig1} also shows that the switching voltages $V_{c,\mathrm{AP}}$ and $V_{c,\mathrm{P}}$ decrease with increasing temperature. The asymmetry of the switching voltages is likely associated with the fringe fields coming from the SAF structure. For this pMTJ we find a bias field of 56~mT at 4~K extracted from field hysteresis loops that favors the P state, lowering $V_{c,\mathrm{AP}}$. Our observations are similar to those of earlier studies~\cite{Cao2019,Lang2020}. 
It is also interesting to note that the pMTJ always switches close to the same resistance values, $R \approx 5849\;\Omega$ for AP$\rightarrow$P and $2530\;\Omega$ for P$\rightarrow$AP transitions. The same behavior was observed in 50 and 60~nm diameter devices.

\begin{figure*}
\includegraphics[width=0.98\textwidth,keepaspectratio]{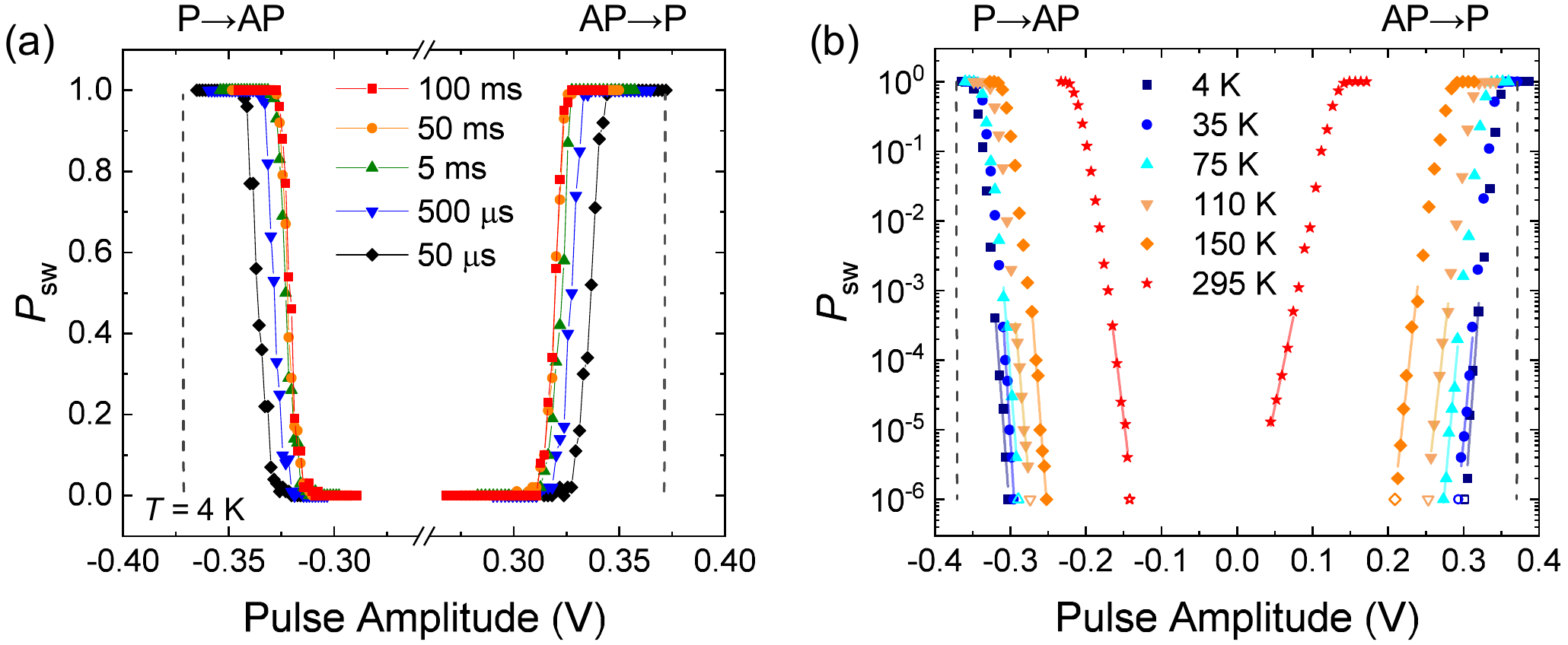}
\caption{(a) Switching probability ($P_\mathrm{sw}$) versus pulsed voltage ramps with various pulse durations of a 40~nm diameter pMTJ at $T_\mathrm{bath}= 4$~K and zero field. Each point is an average of 100 switching trials. (b) Switching probability in the low-voltage (read disturb) limit as a function of voltage pulse amplitudes for the same device at various temperatures. The pulse duration for this measurement was fixed at 10~$\mu$s. The straight lines represent the fits to the data using Eq.~\ref{eq:2} described in the main text. The empty data points at 10\textsuperscript{6} events stand for no errors and were therefore excluded from the fit. The dark gray dashed lines in both panels indicate the behavior expected for a device temperature of 4~K.}
\label{Fig:Fig2}
\end{figure*}
\section{Results and Analysis}
The switching probability was measured and analyzed with the spin-transfer torque switching model to determine the effective switching temperature and threshold voltages $V_\mathrm{c0}$~\cite{Sun2000,Myers2002,Koch2004,LiZhang2004}.
Two methods were used: 1) the determination of switching switching voltage versus pulse duration method at fixed switching probability and 2) measurement of the read disturb rate ($P_\mathrm{sw}\ll 1$) for fixed pulse duration. Both methods have been shown to give reliable estimates of the ratio of the energy barrier to the temperature in magnetic tunnel junctions devices~\cite{Heindl2011}.

In the switching voltage versus pulse duration method the device is set in a know state with a reset pulse and then switching pulses are applied. The state of the pMTJ is measured after each switching pulse, with the junction readout done at a low voltage bias $<15$~mV, much less than the switching voltage. For each set of pulse conditions we study 100 events to determine the switching probability as the number of switching events divided by the total number of events. Figure~\ref{Fig:Fig2}(a) shows the switching probability $P_\mathrm{sw}$ versus pulse amplitude for pulse durations $\tau$ varied over many orders of magnitude, 50~$\mu$s to 100~ms. The results shown in Fig.~\ref{Fig:Fig2}(a) are taken at a temperature of 4~K in zero applied field on the same 40~nm pMTJ as in Fig.~\ref{Fig:Fig1}. See section 1 of the supplementary material~\cite{SM2020} for the experimental results for the 50 and 60~nm devices.

\begin{figure}
\includegraphics[width=0.48\textwidth,keepaspectratio]{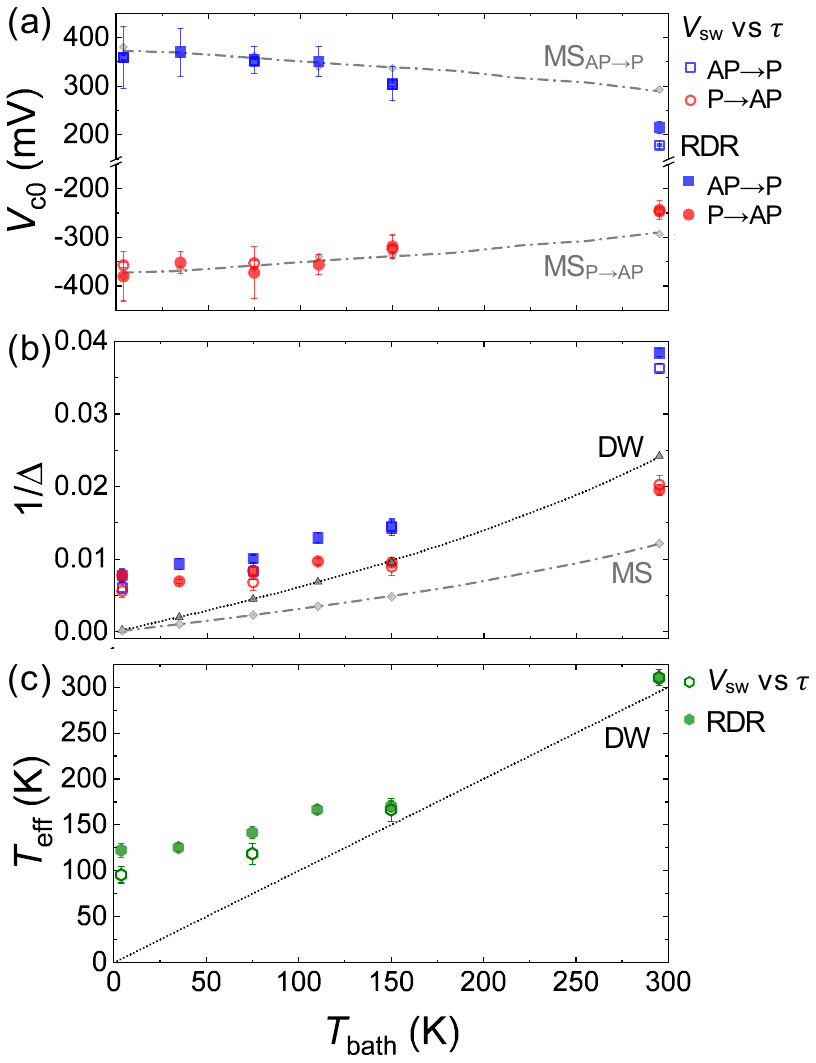}
\caption{(a) Intrinsic switching voltages $V_{c0}$, (b) thermal stability factors $\Delta$ and (c) effective sample temperature $T_\mathrm{eff}$ from the pulsed voltage ramps and RDR measurements at various temperatures between 4 and 295~K. 
The dashed gray and dotted black lines represent the expected behavior for the macrospin (MS) and domain wall (DW) model based on material parameters extracted from VSM and FMR measurements, as discussed in the text.}
\label{Fig:Fig3}
\end{figure}

The thermal activation model relates the switching voltage at fixed switching probability $P_\mathrm{sw}=1-1/e\simeq 0.63$ to the switching temperature $T_\mathrm{eff}$ and $V_{c0}$:
\begin{eqnarray}
V_{sw}&=&V_{c0}\left[1-\frac{1}{\Delta}\ln \left(\frac{\tau}{\tau_0}\right)\right],
\label{eq:1a}\\
\Delta&=&E_\mathrm{b}/k_{B} T_\mathrm{eff}, 
\label{eq:1b}
\end{eqnarray}
where $E_\mathrm{b}$ is the energy barrier to magnetization reversal and $\tau_0$ is the attempt time, $\simeq 1$~ns~\cite{Wernsdorfer1997,Krivorotov2004}. $\Delta$ is the thermal stability factor, the ratio of the energy barrier to the effective switching temperature.
The resulting $V_\mathrm{c0}$ and $\Delta$, obtained by plotting $V_\mathrm{sw}$ versus the logarithm of the pulse duration, are shown in Fig.~\ref{Fig:Fig3}(a) and Fig.~\ref{Fig:Fig3}(b), respectively.

In the read disturb rate (RDR) method the same experimental procedure is applied but lower amplitude write pulses are used, pulses for which the switching probability is very small, $P_\mathrm{sw}\ll1$. We fix the pulse duration $\tau=10\;\mu$s and apply up to a million pulses. This again is done as a function of temperature in zero applied field.  
The resulting probability data is fit to:
\begin{equation}
\ln P_\mathrm{sw}=\ln\left(\frac{\tau}{\tau_0}\right)-\Delta\left(1-\frac{V}{V_{c0}}\right),      
\label{eq:2}
\end{equation}
to determine $V_{c0}$ and $\Delta$.
The fits are shown as straight lines in Fig.~\ref{Fig:Fig2}(b). The hollow data points represent zero errors in $10^6$ switching attempts and are excluded from the fits.
The resulting values for $V_{c0}$ and $\Delta$ obtained by the RDR method can also be found in Fig.~\ref{Fig:Fig3}(a) and Fig.~\ref{Fig:Fig3}(b), respectively. 
$V_\mathrm{c0}$ and 1/$\Delta$ obtained by both methods agree very well (see Fig.~\ref{Fig:Fig3}(a) and (b)). This is expected, as both methods are derived from the same model for thermally activated spin-transfer torque switching~\cite{Sun2000,Myers2002,Koch2004,LiZhang2004}.

We first discuss and analyze the $V_{c0}$ results. $V_{c0}$ begins to saturate at temperatures less than about 150~K, where $V_{c0\mathrm{,4\,K}}$ is 366 and -314~mV and then decreases at room temperature to 215 and -248~mV for AP$\rightarrow$P and P$\rightarrow$AP, respectively (Fig.~\ref{Fig:Fig3}(a)). We also observe that the temperature dependence is slightly bigger for AP$\rightarrow$P then for the P$\rightarrow$AP transition, which is consistent with the data in Fig.~\ref{Fig:Fig1}.

In a macrospin model the intrinsic switching voltage for both switching directions is given by:
\begin{equation}
V_{c0}= \alpha \frac{2e}{\hbar} \left(\frac{1+P^2}{P}\right) \frac{\mu_0 M_{s} H_{k\mathrm{,eff}}\mathcal{V}}{G_\mathrm{P}},     
\label{eq:3}
\end{equation}
for a symmetric junction, a junction in which the materials on both sides of the tunnel barrier have the same spin polarization. Here $\alpha$ is the damping parameter, $e$ is elementary charge, $\hbar$ is the reduced Planck's constant, $\mu _0$ is the the free space permeability, $M_{s}$ is the saturation magnetization, $H_{k\mathrm{,eff}}$ is the effective perpendicular anisotropy, $\mathcal{V}$ is the volume of the free layer, and $G_\mathrm{P}$ is the conductance of the parallel state~\cite{Sun2016}. $P$ is the spin polarization and can be determined from the tunneling magnetoresistance $\mathrm{TMR}= 2P^2/(1-P^2)$~\cite{Julliere1975}. Even in spin torque switching that occurs by nucleation and reversed domain expansion micromagnetic modeling shows that Eq.~\ref{eq:3} accurately characterizes the switching threshold~\cite{Bouquin2018,Mohammadi2020,Volvach2020}.
\begin{table}
\caption{Transport and magnetic properties of the CoFeB/W/CoFeB free layer at selected temperatures. The magnetic properties from FMR and VSM measurements were determined from extended film data as discussed in section 2 of the supplementary materials~\cite{SM2020}.}
\centering
\label{t:Properties}
\setlength{\tabcolsep}{10pt}
\begin{tabular}{lcccc}
\noalign{\smallskip} \hline \hline \noalign{\smallskip}
T & TMR & $M_s t$ & $A_\mathrm{ex}$ & $\mu_0 H_{k\mathrm{,eff}}$ \\
(K) & (\%) & (10\textsuperscript{-3} A) & (pJ m\textsuperscript{-1}) & (T) \\
\hline
4 & 220 & 1.51 & 4.2 & 0.76 \\
35 & 217 & 1.49 & 4.1 & 0.77 \\
75 & 211 & 1.44 & 3.8 & 0.76 \\
110 & 203 & 1.41 & 3.7 & 0.75 \\
150 & 198 & 1.37 & 3.6 & 0.75 \\
295 & 124 & 1.23 & 2.8 & 0.67 \\
\noalign{\smallskip} \hline \hline \noalign{\smallskip}
\end{tabular}
\end{table}

In order to check the correspondence with the expectations of the macrospin model the variation in free layer material parameters with temperature are needed.  We thus measured the magnetic properties of the layer stacks by vibrating sample magnetometry (VSM) and ferromagnetic resonance spectroscopy (FMR) in a field-perpendicular geometry in the temperature range of 4 to 295~K. (See section 2 of the supplementary materials~\cite{SM2020} for the results.) VSM hysteresis loops of the free layer were used to determine its magnetic moment. The resulting magnetic moment per area unit, $M_s t$, where $t$ is the free layer thickness, for different temperatures can be found in Table~\ref{t:Properties}; the values decrease from 1.51 to 1.23~A as the temperature increases, similar but with a stronger dependence on temperature compared to results on dual MgO composite free layers~\cite{Mohammadi2019,Wang2018,Mihajlovic2020}.
Additionally, we also extracted the zero-temperature exchange constant $A_\mathrm{ex}$ for the FL and its dependence on temperature using $A_\mathrm{ex}(T)=A_\mathrm{ex}(0)(M_{s\mathrm{(T)}})/M_{s\mathrm{(0)}})^2$, taking the effective magnetic thickness of this layer to be $t = 1.5$~nm~\cite{Mohammadi2019}. FMR measurements in the field-perpendicular geometry were used to obtain the effective magnetic anisotropy field $H_\mathrm{k,eff}$ of the composite FL and accounts for the demagnetization field associated with its shape~\cite{ChavesOFlynn2015}. A summary of the results is shown in Table~\ref{t:Properties}. We also determined the FL damping parameter from the FMR linewidth versus frequency to be $\alpha$ $\approx$ 0.016 in the whole temperature range.  

We now analyze the results shown in Fig.~\ref{Fig:Fig3}(b), which shows the inverse of the thermal stability factor versus bath temperature. As seen from the definition in Eq.~\ref{eq:1b}, $1/\Delta$ is proportional to the temperature and, therefore, if the effective switching temperature were equal to the bath temperature, $1/\Delta$ would have a zero intercept at zero temperature. This is clearly not the case; $1/\Delta$ is nearly independent of temperature below 75~K. We thus conclude that the effective switching temperature is higher than the bath temperature by at least this amount at 4~K. 

To be more quantitative we use the FL material parameters and the sample geometry to estimate the energy barrier to reversal as a function of temperature. In the macrospin approximation the energy barrier is given by 
\begin{equation}
E_\mathrm{b,MS} = K_\mathrm{eff}\mathcal{V},
\label{eq:5}
\end{equation}
where $K_\mathrm{eff} = \mu_0 M_s H_{k\mathrm{,eff}}/2$ is the effective perpendicular anisotropy. Thus the energy barrier and $V_{c0}$ are dependent on the same temperature dependent material parameters, notably, $M_s$ and $H_{k\mathrm{,eff}}$. The results in Table~\ref{t:Properties} are used to plot the macrospin energy barrier as a function of temperature as the dashed-dotted line in Fig.~\ref{Fig:Fig3}(b). 

It is clear that the macrospin model underestimates 1/$\Delta$ (Fig.~\ref{Fig:Fig3}(b)) for both switching directions, therefore yielding a clear overestimation of $\Delta$ compared to the experimentally obtained values. This is not surprising as the FL is of a size range that we expect the thermally activated reversal to be domain wall mediated. 
Chavez-O'Flynn \textit{et al.}~\cite{ChavesOFlynn2015} estimate the critical diameter above which the reversal is by domain wall motion as $d_\mathrm{c}=(16/\pi) \sqrt{A/K_\mathrm{eff}}\simeq 10$~nm at 4~K, which is much less than the diameter of the FL. In this limit the energy barrier is given by
\begin{equation}
E_\mathrm{b,DW} = 4 \sqrt{A_\mathrm{ex} K_\mathrm{eff}} dt,
\label{eq:6}
\end{equation}
where \textit{d} is the diameter of the nanopillar. The corresponding $1/\Delta$ values are the dotted black lines in Fig.~\ref{Fig:Fig3}(b) and, as expected, are larger than those of the macrospin model. The domain wall model thus gives values of $1/\Delta$ that are closer but still less than the experimental results. 

We use the domain wall model to estimate the effective device switching temperature. First, following common practice, we define $\Delta$ as the average of that for the AP and P states, i.e. $\Delta = (\Delta_\mathrm{AP\rightarrow P} + \Delta_\mathrm{P \rightarrow AP})/2$; we find that the domain wall model describes the data trend between 150 and 295~K well, but still overestimates the values from the measurements below 150~K. 
Assuming this overestimate is associated with heating we extract $T_\mathrm{eff}$ as a function of the cryostat temperature, $T_\mathrm{bath}$. 

The results are shown in Fig.~\ref{Fig:Fig3}(c), with the dotted black line showing $T_\mathrm{eff}=T_\mathrm{bath}$. $T_\mathrm{eff}$ is similar for both the switching voltage and read disturb rate measurement methods. Below 150~K, $T_\mathrm{eff}$ begins to saturates and become independent of the bath temperature.
At $T_\mathrm{bath}$ of 4~K we find a difference between the device and cryostat temperature of $\Delta T_\mathrm{eff} \approx 118$ compared to only 14~K at room temperature in the RDR measurements. The same quantities are 92~K at 4~K and 16~K and 295~K in the switching voltage versus pulse duration method.

\begin{figure}[t]
\includegraphics[width=0.48\textwidth,keepaspectratio]{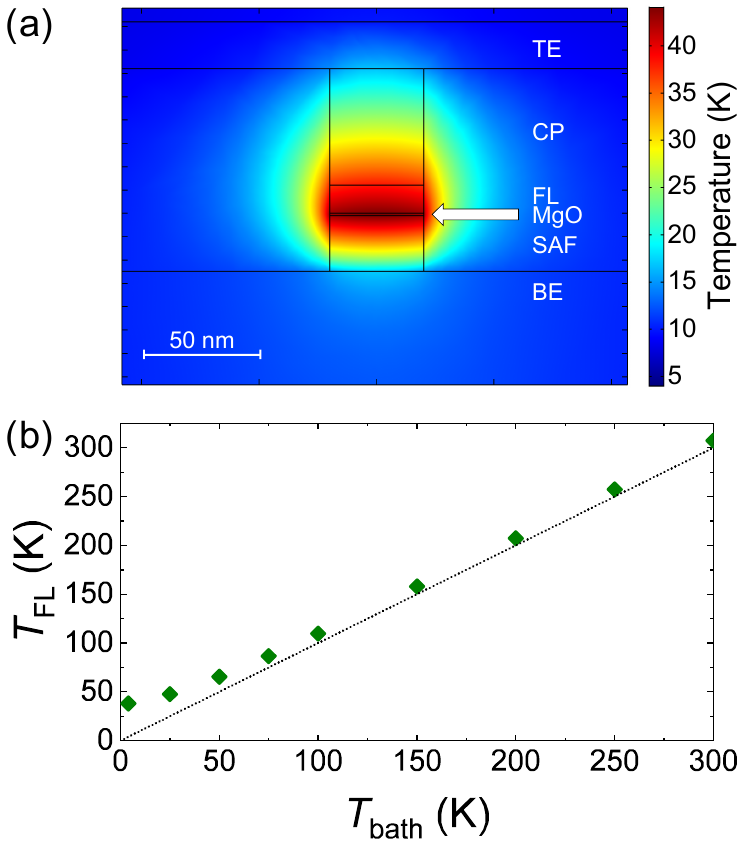}
\caption{COMSOL simulation results. (a) Cross-sectional profile of the thermal map for the simulated 40~nm diameter pMTJ stack at $T_\mathrm{bath}$ = 4~K stressed with 0.3~V and a device resistance of 2.55~k$\Omega$. (b) Free layer temperature $T_\mathrm{FL}$ at various bath temperatures between 4 and 300~K. The dotted black line represents the expected behavior if no heating would occur.}
\label{Fig:Fig4}
\end{figure}
While the temperature rise found at room temperature is consistent with three-dimensional finite-element simulations (COMSOL) of similar MTJ stacks with a composite FL at comparable current densities and RA values~\cite{NamKoong2009}, the saturation of $T_\mathrm{eff}$ towards lower temperatures can not be easily explained.
To gain further understanding of our experimental findings, we performed COMSOL simulations of a 40~nm diameter pMTJ devices at various bath temperatures down to 4~K. The simulations investigate the steady state response of the device since the applied $\mu$s-long pulses are well beyond the time scales for thermalization. We find that the thermal equilibration is reached on the order of nanoseconds, which is also consistent with earlier studies~\cite{Lee2008,NamKoong2009}. Figure~\ref{Fig:Fig4}(a) shows the temperature distribution at the cross-section of our simulated pMTJ device at $T_\mathrm{bath}$ = 4~K. The simulated stack consists of the bottom electrode (BE), three effective tunnel junction layers, SAF, MgO tunnel barrier and FL, a contact plug (capping layer and hard mask, CP), and the adhesion layer with the adjacent top electrode (TE). As expected, the MgO barrier layer is the hottest section of the pillar due to its orders of magnitude lower electrical conductivity compared to the other materials in the stack (315~S/m compared to $10^6-10^7$~S/m for the metal layers). The average temperature we calculated in the free layer at $T_\mathrm{bath}$ = 4~K is 43~K, which is very close to the maximum temperature in the stack. 

Figure~\ref{Fig:Fig4}(b) shows the simulated temperatures in the free layer $T_\mathrm{FL}$ for bath temperatures up to 300~K. We find that our simulation follows the same behavior we observed in our experimental results (Fig.~\ref{Fig:Fig3}(c)). At 300~K we find a small temperature rise $\Delta T$ of around 9~K and it stays close to constant down to 100~K. Below 100~K, we observe significant deviations of $T_\mathrm{FL}$ from the bath temperature (dotted black line in Fig.~\ref{Fig:Fig4}(b)) with $\Delta T$ showing the biggest increase at 4~K ($\Delta T =39$~K). We find that this behavior can be attributed to the fact that the device resistance at the switching point is close to independent of the bath temperature, but the thermal conductivity and heat capacity of our layers decrease with temperature, especially at temperatures below 100~K. We associate the smaller increase in temperature in our simulations compared to the experiment to the idealizations in our model, such as no thermal contact resistance between the layers and encapsulation layers, use of material properties associated with magnetic elements instead of alloys, etc. See section 3 of the supplementary material~\cite{SM2020} for a more detailed description of the simplified stack, the temperature dependent electrical and thermal conductivity as well as the heat capacity of the layers used in our simulation.   

\section{Conclusion}
Our results show that spin-torque switching of pMTJs remains highly probabilistic down to cryogenic temperatures. We associate this with heating of the junction above the bath temperature. Heating becomes more significant as the temperature is reduced as the switching voltage and junction resistance at the switching threshold are nearly independent of temperature but the heat capacity and thermal conductivity of the materials in the junction decrease with decreasing temperature. A further central result of this study is that the switching probability can be described by an effective temperature that becomes independent of the bath temperature at low temperatures. These results are important to furthering the understanding the role of temperature in spin torque switching dynamics of pMTJs and their applications as cryogenic memory.

\begin{acknowledgments}
We thank Jamileh Beik Mohammadi for invaluable support and discussions of the VSM measurements. This research is supported in part by Spin Memory Inc.
\end{acknowledgments}

\bibliography{main}



\end{document}